\RequirePackage{amsmath}
\documentclass{llncs}
\usepackage[utf8]{inputenc}
\usepackage{mathtools}
\usepackage[T1]{fontenc}
\usepackage{amssymb}
\usepackage{bm}
\usepackage[super]{nth}
\usepackage{enumitem}
\usepackage{relsize}
\usepackage{bookmark}
\usepackage{xcolor}
\usepackage{shellesc}
\usepackage{graphicx}%
\usepackage{import}
\usepackage{xargs}
\usepackage{url}
\usepackage[caption=false]{subfig}
\usepackage{thmtools, thm-restate}

\input{svg-macros.tex}

\usepackage{hyperref}
\usepackage[capitalise]{cleveref}
\usepackage[title]{appendix}
\usepackage{microtype}
\usepackage{float}
\usepackage{capt-of}

\definecolor{Qcol}{HTML}{a5c9a5}
\definecolor{Pcol}{HTML}{ff9873}

\usepackage[natbib=true,
   bibstyle=lncs,
   backend=biber,
   hyperref=true,
   url=true,
   isbn=false,
   doi=false,
   eprint=false,
   backref=false,
   citestyle=numeric-comp]{biblatex}

\input{macros.tex}

\title{Technical Report: Property-Directed Verified Monitoring of Signal~Temporal~Logic}

\author{Thomas Wright\orcidID{0000-0001-8035-0884} \and Ian Stark\orcidID{0000-0001-6800-812X}}

\institute{
School of Informatics,
University of Edinburgh, UK}

\begin{document}

\maketitle
\begin{abstract}
Signal Temporal Logic monitoring over numerical simulation traces has emerged as an effective approach to approximate verification of continuous and hybrid systems. In this report we explore an exact verification procedure for STL properties based on monitoring verified traces in the form of Taylor model flowpipes as produced by the \emph{Flow*} verified integrator. We explore how tight integration with Flow*'s symbolic flowpipe representation can lead to more precise and more efficient monitoring. We then show how the performance of monitoring can be increased substantially by introducing masks, a property-directed refinement of our method which restricts flowpipe monitoring to the time regions relevant to the overall truth of a complex proposition. Finally, we apply our implementation of these methods to verifying properties of a challenging continuous system, evaluating the impact of each aspect of our procedure on monitoring performance.
\end{abstract}

\section{Introduction}

Signal Temporal Logic (STL)~\cite{maler2004monitoring} is an established and effective framework for describing and monitoring temporal properties of real-valued signals in continuous time, enabling verification of both continuous and hybrid systems.
Much work on STL has focused on monitoring signals derived from numerical simulation traces.  This is a powerful technique, but the approximate nature of such traces can lead to erroneous signal values at some timepoints whilst the signals do not account for uncertainties in the underlying model.
Recently \citet*{ishii2016intervalmonitoring} have explored combining signal-based monitoring techniques with interval analysis to perform exact verification of STL properties over traces produced via verified integration.

In this report we propose a new verified STL monitoring algorithm based on preconditioned Taylor model flowpipes~\cite{makino2011tmpreconditioning} generated with the \emph{Flow*} verified integrator~\cite{chen2013flowstar}.
Our algorithm starts from such a Flow* flowpipe, which tracks our uncertain knowledge of the system state at each point in time (whether due to numerical errors or uncertain model parameters), and produces a \emph{three-valued signal} reflecting our resulting uncertain knowledge in the truth value of a STL property over time.
One of Flow*'s key features is its sophisticated symbolic flowpipe representation, which allows it to handle a wide range of non-linear continuous and hybrid systems~\cite{chen2015benchmark}, but poses challenges to effectively monitoring properties over these flowpipes, since overapproximating the value of the flowpipe at a given timepoint can be expensive and requires a careful tradeoff between accuracy and efficiency.
We tackle these challenges by tightly integrating our monitoring strategy for atomic propositions with Flow*'s flowpipe representation, in contrast to most previous verified monitoring approaches which treat the flowpipe as a closed box by evaluating it as an interval function.
This allows us to vary our evaluation strategy on demand at each timepoint, 
as required to verify a given atomic proposition.
We are thus able to maximize precision for complex atomic propositions by utilizing Taylor model arithmetic to avoid the \emph{dependency problem}~\cite{berz1998tm},
or fall back to simpler interval evaluation strategies when this suffices to determine the value of the signal at the current timepoint.

We further refine our method by introducing \emph{masks},
a special type of signal representing the region of time for which each proposition of a STL formula is required by the overall monitoring process.
We present methods for computing masks in a top-down manner, complementing the normal bottom-up signal monitoring process.
We also see how to efficiently monitor each atomic proposition under a given mask.
This allows us to handle the monitoring of each atomic proposition as a single-pass offline verification problem with a mask providing a condensed view of the regions of interest; we are hence able to utilize Taylor model methods to refine the signal in the most crucial time regions whilst avoiding unnecessary work elsewhere.
Altogether,
this gives a property-directed algorithm for efficient and precise STL monitoring over Flow* flowpipes.

The structure of this report is as follows.
In \cref{sec:related-work} we review related work.
\cref{sec:background} covers necessary background information regarding interval arithmetic, Taylor Models, Flow*, and STL, and establishes our notation.
In \cref{sec:signals} we introduce our verified monitoring algorithm by defining three-valued signals and our method of monitoring atomic propositions over Flow* flowpipes.
In \cref{sec:masks} we present masks and show how they may be used to perform property-directed monitoring.
In \cref{sec:evaluation} we evaluate the benefits of each part of our method for monitoring STL properties in a complex continuous system.
Finally, in \cref{sec:conclusion} we present our conclusions and discusses future directions.

\subsection{Related Work}
\label{sec:related-work}

The most closely related work is the verified STL monitoring algorithm of Ishii~et~al.~\cite{ishii2015ltl,ishii2016intervalmonitoring,ishii2017hysia}.
We build upon their approach of monitoring atomic propositions using interval root finding to derive verified signals; however, whilst their method treats the result of verified integration as a generic interval function, we work directly with the symbolic Taylor model flowpipes produced by Flow*.
Our formulation of verified signals is also quite different, representing signals using three-valued logic rather than inner and outer interval approximations.
Other works have looked at three-valued extensions of STL~\cite{banks2017moresensitivecontext,vissat2017tstl}, but have not applied this in the context of exact, interval-based, formal verification of continuous systems. Three valued logic has also been applied to model checking LTL properties over discrete systems with partially known state spaces~\cite{bruns1999threevaluedltl}. \citet{fisman2018trincompletepaths}~also investigated a more general semantics for Linear Temporal Logic (LTL) and STL over a number of different forms of incomplete traces.

A variety of different approaches have been explored for formal verification of temporal logics over hybrid and continuous systems.
Early methods include \cite{alur1996automatic} which initially focused on linear systems and the temporal logic TCTL.
\citet{piazza2005algorithmic} developed methods for temporal logic verification of semi-algebraic hybrid systems based on quantifier elimination.
\citet{bresolin2013hyltl} developed a method for encoding LTL properties of hybrid systems as hybrid automata reachability problems, extending the common automata theoretic approach~\cite{vardi1986automata} to temporal logic verification, allowing existing reachability tools to be applied to LTL model checking. However, applying this method to even relatively simple LTL properties can result in large hybrid automata which are beyond the abilities of current reachability analysis tools.
\citet{cimatti2014kliveness} presented another approach to reducing LTL verification on hybrid systems to reachability analysis.
The dynamic temporal logic dTL\textsuperscript{2}~\cite{jeannin2014dtl2} takes a rather different approach, including both hybrid systems and nested temporal modalities as part of the logic and providing an associated proof calculus.

A number of recent works have focused specifically on exact STL verification using ideas from reachability analysis. 
\citet{roehm2016reachset} provided an approach to STL verification for hybrid automata based on checking (discrete) time sampled versions of STL formulae against reachsets produced by reachability analysis tools such as CORA~\cite{althoff2015CORA}.
\citet{bae2019syntacticseparation} introduced an approach to STL verification which translates properties into constraint problems which can be verified exactly using a SMT solver such as Z3 for linear systems~\cite{de2008z3} or the $\epsilon$-complete decision procedure dReal for non-linear systems~\cite{gao2013dreal,kong2015dreach}.
This work is the closest to providing an automated property-directed model checking procedure for STL. The constraint solving tools on which this work relies are very different from verified integration tools such as Flow*, and currently which approach performs better depends heavily on the system at hand~\cite{chen2015benchmark,liu2015variabletransfromation}. These exact verification-based methods also compete with approximate methods such as~\cite{donze2010breach,fainekos2007robustsamplingMITL,annpureddy2011staliro,banks2017moresensitivecontext} which attempt to verify properties by sampling trajectories and may use quantiative measures of robust satisfaction~\cite{donze2010robustness,fages2008biochamrobustness} to give some assurance of robustness of results and to cover uncertain model parameters and initial conditions.

Our method has some similarities with online monitoring in that in both cases we have partial knowledge of the underlying signal, and wish to avoid unnecessary (re)evaluation of signals for atomic propositions. \citet{deshmukh2017robustintervalstl} introduced an interval quantitative semantics for STL in order to perform online monitoring over partial traces. In their method partiality of traces is used to record the fact that part of the trace has not yet been received at the time of monitoring whereas in our method it reflects uncertainty in the underlying system. Both~\cite{deshmukh2017robustintervalstl} and the earlier incremental marking procedure~\cite{nickovic2007amt,maler2008stlcheckingbookchapter} attempt to avoid evaluation of atomic propositions at timepoints unnecessary for the operators in which they occur similarly to masks, however, we statically calculate a mask for the timepoints of interest in a complete proposition, whilst online monitoring algorithms calculate (contiguous) time horizons for propositions each time they are revisited. Masks also play a quite different role to these optimizations, since masks are computed in a top-down manner to allow specific parts of the verified monitoring process to be avoided on a non-contiguous region throughout the time domain, whilst the role of these optimizations in online monitoring is to reduce memory usage by allowing time points outside of a proposition's time horizon to be forgotton or to allow early termination of simulations.

\section{Background}
\label{sec:background}

In this section we review some background material including interval arithmetic and Taylor models, which will form the centre of our monitoring process.

\subsection{Interval Arithmetic}

We will work with interval arithmetic~\cite{moore2009intervalanalysis} over closed intervals 
and denote the lower and upper endpoint of interval
\(
    I
    = \interval{l_I}{u_I}
    \triangleq \bigl\{ x \in \mathbb R
              \mathrel{\big|}
              l_I \leq x \leq u_I \bigr\}
\) 
by
\(
    l_I, u_I \in \mathbb R \cup \{-\infty, \infty\}
\)
respectively.
Arithmetic operations can be computed using interval endpoints so that
\begin{alignat*}{6}
    &&I + J & {}\triangleq{} && \bigl\{ x + y
                 &&\mathrel{\big|}
              x \in I, y \in J \bigr\}
    & {}={} & \interval{l_I + l_J}{{}&&u_I + u_J}
\end{alignat*}
and $IJ$ and $I - J$ are computed similarly. We can also define set operations on intervals so
\( 
    I \cap J = \interval{\max\{l_I, l_J\}}{\min\{u_I, u_J\}},
\)
and whilst the set-theoretic union $I \cup J$ is not necessarily an interval, we may over-approximate and use
\(
    I \cup J
    \triangleq
    \interval{\min\{l_I, l_J\}}{\max\{u_I, u_J\}}
\)
in its place.

Given a real-valued function $f : I \to \mathbb R$ over interval domain $I$,
an interval valued function $F$ is an \emph{interval extension of $f$} if $f(x) \in F(X)$ for every point $x$ and interval $X$ such that $x \in X$.
Assuming $f$ is differentiable and we have an interval extension $F'$ of its derivative $f'$,
we may apply the \emph{Extended Interval Newton method}, a generalisation of the Newton root finding method, which uses interval evaluation of $F$ to produce guaranteed interval enclosures of all roots of $f$~\cites{,ishii2016intervalmonitoring}[Chapter 8.1]{moore2009intervalanalysis};
we denote the set of such intervals as $\operatorname{roots}(F, F', I)$.

\subsection{Taylor Model Arithmetic}

When used in verified numerical computation, interval arithmetic often leads to imprecise or inconclusive results due to the so called \emph{dependency problem} in which the functional dependencies within an expression are ignored by an interval overapproximation, causing the approximation errors to be compounded throughout the computation.
This motivates the use of symbolic methods such as Taylor models~\cite{makino1999dependency,berz1998tm}, which give a higher-order symbolic over-approximation of a function based on a Taylor series expansion and an interval remainder bound.
\begin{definition}
    Given a function \(\mathbf f:\mathbf D \to \mathbb R^n\) with domain \(\mathbf D \subseteq \mathbb R^m\), a \emph{\kth-order Taylor model for~$\mathbf f$} is a pair \((\mathbf p, \mathbf I)\) of an $n$-dimensional vector \(\mathbf p\) of \kth-order $m$-variable polynomials and an $n$-dimensional box~$\mathbf I$ such that
    \(
        f(\mathbf x) \in \mathbf p(\mathbf x) + \mathbf I
    \)
    for all \(\mathbf x \in \mathbf D\). 
\end{definition}
The precision of the approximation increases as the order of the Taylor model increases and the size of the remainder decreases.
Given two Taylor models \((\mathbf p, \mathbf I)\) and \((\mathbf q, \mathbf J)\) enclosing functions $\mathbf f$ and~$\mathbf g$ respectively, we can carry out various operations symbolically.
Addition of Taylor models may be defined piecewise, so that if \(f\) has Taylor model \((\mathbf p_f, \mathbf I_f)\) and \(g\) has Taylor model \((\mathbf p_g, \mathbf I_g)\) (both of order \(k\)) then \(f + g\) has Taylor model
\[
    (\mathbf p_f, \mathbf I_f) + (\mathbf p_g, \mathbf I_g) =
    (\mathbf p_f + \mathbf p_g, \mathbf I_f + \mathbf I_g) .
\]

Multiplication of Taylor models is similar except, in order to preserve the order of the Taylor model, we must truncate higher-order terms into the remainder interval.
Hence we define a \kth-order Taylor model for \(\mathbf f \mathbf g\) over domain \(\mathbf D\) by
\[
    (\mathbf p_f, \mathbf I_f) (\mathbf p_g, \mathbf I_g) =
    (\mathbf p_f \mathbf p_g - \mathbf p_E, \mathbf I_f \mathbf g(\mathbf D) + \mathbf f(\mathbf D) \mathbf I_g + \mathbf I_f \mathbf I_g + \mathbf p_E(\mathbf D)) .
\]
where \(\mathbf p_E\) contains all terms from \(\mathbf p_f \mathbf p_g\) of order strictly greater than \(k\).
Other common operations including multiplication, division, and differentiation can be carried out with a suitable expansion of the remainder intervals~\cite{berz1998tm}.

Finally, we may define the \emph{symbolic composition} 
\(
    (\mathbf q, \mathbf J)
    \mathop{\square}
    {(\mathbf p, \mathbf I)}
    ,
\)
by symbolically substituting each dimension of $(\mathbf p, \mathbf I)$ into $(\mathbf q, \mathbf J)$ in place of the corresponding variable. This gives a Taylor model which is guaranteed to enclose the composition~$\mathbf g \circ \mathbf f$ of the underlying functions~\cite{berz1998tmcomputation}. 

\subsection{Flow* Verified Integration}

Consider a continuous system described by an $n$-dimensional collection of ODEs
\begin{align}
    \label{eq:continuous-system}
    \frac{\mathrm d\mathbf{x}}{\mathrm d t}
    = \mathbf f(\mathbf x, t)
\end{align}
with initial conditions~$\mathbf x(0)$ in some starting set $S \subseteq \mathbb R^n$ and with $\mathbf f$ being, for example, Lipschitz continuous.  Such a system has a unique solution $\mathbf x_{s}$ for any given initial condition $\mathbf s \in S$~\cite{kolmogorov1975realanalysis}.
The aim of \emph{verified integration} is to find an interval function $\mathbf F:\intervalinline{0}{T} \to \mathbb R^n$ enclosing every solution~$\mathbf x_s$ for $s \in S$ over a bounded time window $\intervalinline{0}{T}$.
Flow*~\cite{chen2013flowstar} uses verified integration based on Taylor models~\cite{berz1998tm} to compute such an enclosure covering the whole set $S$, soundly accounting for the uncertainty in the initial point $s$ as well as floating point rounding errors.
Flow* represents the solution as a \emph{flowpipe} consisting of a sequence of \emph{preconditioned Taylor models}~\cite{makino2011tmpreconditioning}, that is, over each interval time step $\intervalinline{t_k}{t_{k+1}} \subseteq \intervalinline{0}{T}$, the solution is enclosed in a composition of two Taylor models
\[
    \left(\mathbf p^{(k)}, \mathbf I^{(k)}\right)
    \triangleq
    \left(\mathbf p_{\text{post}}^{(k)},
    \mathbf I_{\text{post}}^{(k)}\right)
    \square
    \left(\mathbf p_{\text{pre}}^{(k)},
    \mathbf I_{\text{pre}}^{(k)}\right)
    .
\]
Whilst this representation is extremely powerful, allowing Flow* to handle nonlinear systems with complex continuous dynamics, it can be expensive to work with the generated flowpipes: each pair of preconditioned Taylor models must be composed symbolically (and then be placed into Horner form~\cite{pena2000multivariatehorner,chen2015thesis}) in order to carry out accurate interval evaluation.
This step is a prerequisite for applying most forms of analysis to the flowpipe including reach-avoidance checking, plotting~\cite{chen2013flowstar}, and the verified monitoring algorithm~\cite{ishii2015ltl}.

\subsection{Signal Temporal Logic}

Signal Temporal Logic~\cite{maler2004monitoring} specifies properties of continuous trajectories in dynamical systems such as \cref{eq:continuous-system}. Propositions are defined according to the grammar,
\[
    \phi, \psi
    \bnfdef 
    \rho
    \bnfbar
    \phi \wedge \psi
    \bnfbar
    \phi \vee \psi
    \bnfbar
    \neg \phi
    \bnfbar
    \phi \UI{I} \psi\;\text{,}
\]
where $I$ is an interval and atomic propositions consist of inequalities $\rho \triangleq p > 0$ defined by polynomials $p$ over the system's variables.

The semantics for the connectives are defined by specifying when a system trajectory $\mathbf x$ satisfies a property $\phi$ at time $t$, written $(\mathbf x, t) \models \phi$.
Then the semantics for atomic propositions is given by $(\mathbf x, t) \models p > 0$ iff $p(\mathbf x(t)) > 0$ and the normal boolean connectives are extended to each time point so that, for example, $(\mathbf x, t) \models \phi \wedge \psi$ iff $(\mathbf x, t) \models \phi$ and $(\mathbf x, t) \models \psi$.
The fundamental temporal modality \emph{until} is defined by $(\mathbf x, t) \models \phi \UI{I} \psi$ iff there exists $t' \in t + I$, such that $(\mathbf x, t') \models \psi$ and for all $t'' \in [t, t']$, $(\mathbf x, t'') \models \phi$.
We can also define the derived temporal modalities $\FI{I}(\phi) \equiv \true \UI{I} \phi$~\emph{(eventually)} and $\GI{I}(\phi) \equiv \neg \FI{I}(\neg \phi)$~\emph{(globally)}.

The problem of \emph{verified monitoring} of a STL property $\phi$ at a time point $t \in [0, T]$ is to determine whether or not $(\mathbf x_s, t) \models \phi$ for every trajectory $\mathbf x_{s}$ of a given system.

\section{Three-Valued Monitoring over Flow* flowpipes}
\label{sec:signals}

In this section we present our basic verified monitoring algorithm for Signal Temporal Logic.
In \cref{sec:tvstl} we introduce three-valued signals and specify rules to combine these to derive signals for complex propositions.
We then develop an efficient algorithm for monitoring signals of atomic propositions based over Flow* flowpipes in \cref{sec:signals-for-atomics,sec:efficient-taylor-model-monitoring}.

\subsection{Three-Valued Signals}
\label{sec:tvstl}

Our verified monitoring algorithm is based on \emph{three-valued signals}:
\begin{definition}
    A \emph{three-valued signal} is a function $s: [0, \infty) \to \{\true, \unknown, \false\}$.
\end{definition}
These extend the boolean signals $s:[0, \infty) \to \{\true, \false\}$ used in numerical STL monitoring algorithms to track the validity of the answer at each time point $t \in [0, \infty)$, to allow a third answer, \emph{Unknown} ($\unknown$), if we can neither verify nor refute the proposition.
We interpret these logic values under the rules of Kleenian three-valued logic so $\false \vee \unknown \equiv \unknown \equiv \true \wedge \unknown$.
A three-valued signal is consistient with the (boolean) truth values of a proposition $\phi$ if it soundly under-approximates the time-regions on which $\phi$ is True and False:
\begin{definition}
    Given a proposition $\phi$ and a three-valued signal~$s$, we say \emph{$s$ is a signal for~$\phi$}  (over the trajectories of a given system) if at every time~$t$,
    \begin{alignat*}{3}
        s(t) &= \true &\qquad&\implies\qquad& (\mathbf x, t) \models \phi &\quad \text{for every trajectory $\mathbf x$} \\
        s(t) &= \false &\qquad&\implies\qquad& (\mathbf x, t) \not\models \phi &\quad \text{for every trajectory $\mathbf x$}\;\text{.}
    \end{alignat*}
\end{definition}
This definition allows a single proposition $\phi$ to be approximated by many different signals to differing degrees of precision. Indeed, the signal which is unknown everywhere is a valid (but uninformative) signal for every proposition.

For concrete computation we work with those three-valued signals~$s$ that can be represented by a finite set of nonempty disjoint intervals $I_j = [a_j, b_j]$ and logical values $s_j \in \{\true, \false\}$, with $j$ ranging over some index set~$\Gamma$.  These values determine signal~$s$ on those intervals, with $\unknown$ assumed elsewhere, and we write $s=(I_j,s_j)_j$.  For convenience we also admit some improper representations including empty intervals, values $s_j = \unknown$, and intervals with overlapping endpoints (but consistent logical values); these may all be rewritten as proper representations. 

Given propositions $\phi$ and~$\psi$ with $s = (I_j, s_j)_j$ a signal for~$\phi$ and $w = (I_j, w_j)_j$ a signal for~$\psi$, we have the following constructions:
\begin{description}[align=right,labelwidth=4em]
    \item[$\neg \varphi$] has a signal given by $\neg s = (I_j, \neg s_j)_j$
    \item[$\varphi \wedge \psi$] has a signal given by $s \wedge w = (I_j, s_j \wedge w_j)_j$
    \item[$\FI{[a,b]} \varphi$] has a signal given by $\FI{[a,b]} s = (K_j \cap [0, \infty), s_j)_j$
    where %
    \[
        K_j = \begin{cases}
            
            I_j - [a,b]
            & \text{if }s_j = \true \\
            (I_j - a) \cap (I_j - b)
            & \text{if }s_j = \false
        \end{cases}
    \]
\end{description}
In the above we have assumed, without loss of generality, that $s$ and~$w$ are represented using a common set of intervals $I_j$; this is always possible by taking a common refinement of the representations of $s$ and~$w$ respectively.

In the case of boolean signals, the until signal $s \UI{J} w$ is usually computed by subdividing $s$ into \emph{disjoint unitary signals} $s_{j}$
(that is, signals which are indicator functions $s_j(t) = \indicator{I_j}(t) = \text{($\true$ if $t \in I_j$ otherwise $\false$)}$ of pairwise disjoint intervals)~\cite{maler2004monitoring}.
For three-valued signals we will follow a similar approach, however we need an appropriate three-valued generalisation of unitary signals. To this end we define a \emph{connected signal}.
\begin{definition}
    We say a three-valued signal~$s$ is \emph{connected} if for every interval $[a, b]$ we have that,
    \begin{alignat*}{3}
        s(a) \wedge s(b) \leq s(t)
        &&\qquad&&\text{for all $t \in [a, b]$}.
     \end{alignat*}
     under the ordering $\false \lneq \unknown \lneq \true$.
\end{definition}

\begin{proposition}
    A three-valued signal $s$ is connected iff there exist intervals $J \subseteq I$ such that $s$ is equal to the \emph{three-valued indicator signal},
    \[
        \tvindicator{J}{I}(t)
        \triangleq
        \begin{cases}
            \true & \text{if $t \in J$} \\
            \unknown & \text{if $t \in I \setminus J$} \\
            \false & \text{if $t \not\in I$}
        \end{cases}
    \]
\end{proposition}
\begin{proof}
	First we note that an indicator signal~$\tvindicator{J}{I}$ is clearly connected. Conversely, given a connected signal~$s$, take
	\begin{alignat*}{4}
		J & ={}& \bigg[\;
			j_l & \triangleq \inf_{s(t) = \true} t,\,&
			j_u & \triangleq \sup_{s(t) = \true} t&\bigg]&
			\quad \text{ and}\\
		I & ={}& \bigg[\;
			i_l & \triangleq \inf_{s(t) \in \{\true,\unknown\}} t,\,\quad&
			j_u & \triangleq \sup_{s(t) \in \{\true, \unknown\}} t&\;\bigg]& \supseteq J
	\end{alignat*}
	(for simplicity we assume that the properties defining these intervals hold at their endpoints; the other cases result in open/half-open intervals and are analogous). Then we see that $s = \tvindicator{J}{I}$: for any $t \in J$,  $\true \geq s(t) \geq \min(s(j_l), s(j_u)) = \true$ by the connectedness of $s$, whilst for any $t \not\in I$, supposing $s(t) \neq \false$ and w.l.o.g. that $t > j_u$, since $s$ is connected, $s([t_u, t]) = \{\true, \unknown\}$ contrary to the maximality of $j_u$, and finally, for any $t \in I \setminus J$, we first see by connectedness of $s$ that $s(t) \in \{\true, \unknown\}$, and then if we suppose $s(t) = \true$ we get a contradiction, similarly to the previous case, leading us to conclude that $s(t) = \unknown$. 
\end{proof}

We note that it is straightforward to compute a signal for $\phi \UI{K} \psi$ on connected signals.
\begin{proposition}
    If $s$ and $w$ are respectively signals for $\phi$ and $\psi$, and $s$ is connected then
    \[
        s \UI{K} w \triangleq s \wedge \FI{K} \!\left(
            s \wedge w
        \right)
    \]
    is a signal for $\phi \UI{K} \psi$.
\end{proposition}
\begin{proof}
        Suppose $(s \UI{K}\!\! w)(t) = \true$.
        Then $s(t) = \true$ and for some $t' \in t + K$, $s(t') = \true$ and $w(t') = \true$.
        But then since $s$ is connected, $s(t') = \true$ for all $t^{\prime\prime} \in [t, t']$, showing that $(\mathbf x, t) \models \phi \UI{K} \psi$. 

        Suppose $(s \UI{K}\! w)(t) = \false$.
        Then either $s(t) = \false$ in which case $(\mathbf x, t) \not\models \phi \UI{K} \psi$, or for all $t' \in t + K$, $s(t) = \false$ or $w(t) = \false$, in which case again $(\mathbf x, t) \not\models \phi \UI{K} \psi$.
    \end{proof}

We next decompose a three-valued signal into connected signals. 

\begin{proposition}
    Given a three-valued signal $s$ and disjoint intervals $I_j$ such that $s^{-1}(\{\true,\unknown\}) = \biguplus_j I_j$, we have a decomposition
    \(
        s = \bigvee_j \bigvee_k s_{j,k}
    \)
    of $s$ into the connected components:
    \begin{itemize}
        \item $s_{j,0} = \tvindicator{\varnothing}{I_j}$
        whenever $I_j \cap s^{-1}(\{\true\}) = \varnothing$;
        \item $s_{j,k} = \tvindicator{J_{j,k}}{I_j}$
            given intervals $J_{j,k}$ such that 
           $I_j \cap s^{-1}(\{\true\}) = \biguplus_k J_{j,k}$.
    \end{itemize}
\end{proposition}
\begin{proof}
	Give any time $t$, if $s(t) = \false$ then for any $I_j$ and any $J \subseteq I_j$ since $t \notin s^{-1}(\{\true, \false\}) = \biguplus_j I_j$.

	If  $s(t) = \true$, then we see that $t \in J_{j,k} \subseteq I_j$ for some $j,k$. But then
	\[
		\true
		\geq \bigvee_l \bigvee_m s_{l,m}(t)
		\geq s_{j,k}(t)
		= \true.
	\]

	If $s(t) = \unknown$ then we have $t \in I_j$ for some unique $j$ and so
	\[
		\bigvee_l\bigvee_m s_{l,m}(t) = 
		\bigvee_m s_{j,m}(t).
	\] 
	But then we note $s_{j,0}(t) = \tvindicator{\varnothing}{I_j}(t) = \unknown$ and for $m > 0$, $s_{j,m}(t) = \tvindicator{J_{j,k}}{I_j}(t) = \unknown$, showing that the RHS is $\unknown$ also.
\end{proof}

\begin{figure}
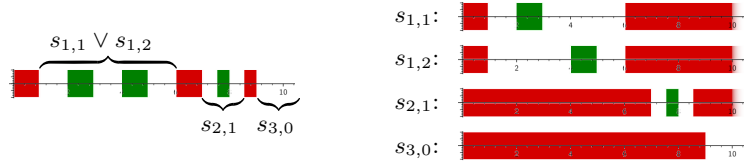

    \centering
    \myincludesvg[0.8\textwidth]{decomposed-monitoring}
    \caption{The decomposition of $s$ into components $s \equiv (s_{1,1} \vee s_{1,2}) \vee s_{2,1} \vee s_{3,0}$.}
    \label{fig:signal-decomposition}
\end{figure}

\begin{example}
    \label{ex:three-valued-decomposition}
    Given the three-valued signal
    \[
        s = (([0, 1], \false), ([2, 3], \true), ([4, 5], \true),
               ([6, 7], \false), ([7.5, 8], \true), ([8.5, 9], \false))
    \]
    we have the decomposition~(\cref{fig:signal-decomposition})
    \[
        s = (s_{1,1} \vee s_{1,2}) \vee 
            s_{2,1} \vee
            s_{3,0}
          = (\tvindicator{[2,3]}{(1,6)} \vee \tvindicator{[4,5]}{(1,6)}) \vee
            \tvindicator{[7.5, 8]}{(7, 8.5)} \vee
            \tvindicator{\varnothing}{(9, \infty)}.
    \]
\end{example}

We now use this decomposition to construct a signal for $\phi \UI{K} \psi$:
\begin{proposition}
    \label{prop:U-decomposition}
    If $\phi$ has a three-valued signal $s = \bigvee_j\bigvee_k s_{j,k}$ with connected components $s_{j,k}$ and $\psi$ has a signal $w$, then $\phi \UI{K} \psi$ has a signal given by,
    \begin{align}
        \label{eq:until-decomposition}
        s \UI{K} w
        \triangleq
        \bigvee_j\bigvee_k
        s_{j,k} \wedge \FI{K}\!\left(
           s_{j,k} \wedge w
        \right).
    \end{align}
\end{proposition}
\begin{proof}
	If $(s \UI{K} w)(t) = \true$ then for some $j, k$,
	\[
		s_{j,k} \wedge \FI{K}\!\left(s_{j,k} \wedge w\right)(t) = \true.
	\]
	Hence $s_{j,k}(t) = \true$ and for some $t' \in t + K$, $s_{j,k}(t') = \true$ and $w(t) = \true$.
	However, since $s_{j,k}$ is connected, we conclude $s_{j,k}(t'') = \true$ for all $t'' \in [t, t']$ and hence $(\mathbf x, t)\models \phi$ for all $t'' \in [t, t']$, showing $(\mathbf x, t) \models \phi \UI{K} \psi$.

	Suppose that $(\mathbf x, t) \not\models \phi \UI{K} \psi$.
	Then $s(t) \neq \false$ and for some $t' \in t + K$, $w(t) \neq \false$ and for all $t'' \in [t, t']$, $s(t) \neq \false$ and $w(t) \neq \false$.
	Then since $s(t'') \neq F$ on the connected region $[t, t']$, there must hence be some $l$ with $[t, t'] \subseteq I_m$.
	But then, (picking some $m$ given $l$),
	$s_{lm} \wedge \FI{K}\!\left(s_{lm} \wedge s_\psi\right)(t) \neq \false$, and so
	\[
		(s \UI{K} w)(t)
		=
		\bigvee_j \bigvee_k s_{j,k} \wedge \FI{K}\!\left(s_{j,k} \wedge w\right)(t)
		\neq \false
	\]
	completing the proof by contrapositive.
\end{proof}

\subsection{Signals for Atomic Propositions}

\label{sec:signals-for-atomics}

\begin{figure}[b]
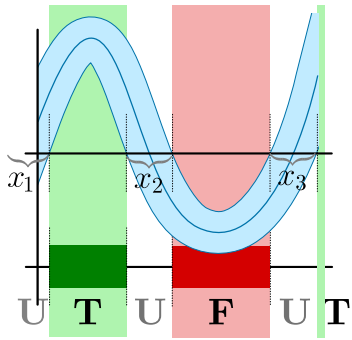

    \centering
    \myincludesvg[14em]{monitoring-atomic}
    \caption{Transition from root~finding to three-valued~signals.}
    \label{fig:monitoring-atomic}
\end{figure}

We now turn our attention to generating a signal for an atomic proposition $\rho \triangleq p > 0$ with defining polynomial $p$, based on a Flow* flowpipe.

We do this in a single pass algorithm which iterates over each time segment of the flowpipe and encloses all roots of $p$ for the current timestep.
For each timestep $[t_k, t_{k+1}]$, Flow* provides two Taylor models, 
\(
    \left(
        \mathbf q_{\text{post}}^{(k)},
        \mathbf I_{\text{post}}^{(k)}
    \right)
\)
and
\(
    \left(
        \mathbf q_{\text{pre}}^{(k)},
        \mathbf I_{\text{pre}}^{(k)}
    \right)
\)
whose composition encloses all trajectories of the system over the time step.
The value of $p$ over system trajectories is hence enclosed by the Taylor model $G_p^{(k)}(\mathbf s, t)$ defined by the composition
\begin{equation}
    \label{fig:p-taylor-model}
    G_p^{(k)}
    \triangleq
    p
    \fpcomp
    \left(
        \mathbf q_{\text{post}}^{(k)},\,
        \mathbf I_{\text{post}}^{(k)}
    \right)
    \fpcomp
    \left(
        \mathbf q_{\text{pre}}^{(k)},\,
        \mathbf I_{\text{pre}}^{(k)}
    \right)
\end{equation}
where $t \in [t_k, t_{k+1}]$ and $\mathbf s$ ranges over the $n$-dimensional box $[-1, 1]^n$~\cite{chen2015thesis,makino2011tmpreconditioning}.
Therefore, we have an interval extension of $p$ over the time interval $[t_k, t_{k+1}]$ given by the interval function
\(
    H_p^{(k)}(t) \triangleq G_p^{(k)}([-1, 1]^n, t)
\)
which may be evaluated using interval arithmetic.

We may then determine a signal for $\rho$.
\begin{proposition}
    Given atomic proposition $\rho = p(\mathbf x) > 0$, the three-valued signal 
    \(
        s \triangleq (I_j, s_j)_j
    \)
    is a signal for $\rho$ where $I_j$ are the interval components of \[
        [0, T] \setminus \bigcup \left\{
            x_0
            \;\Big|\;
            x_0 \in \operatorname{roots}\!\left(
                H_p^{(k)},
                \frac{\mathrm d}{\mathrm d t}H_p^{(k)},
                [t_k, t_{k+1}]
            \right)
            \text{for some $k$}
        \right\} 
        .
    \]
    and $s_j$ is $\true$ iff $H^{(k)}_p(t') > 0$ for some $k$ and $t' \in I_j\cap[t_k, t_{k+1}]$. 
\end{proposition}
The unknown regions are given by amalgamating the roots of $H^{(k)}_p$ over each time step.
These roots are soundly enclosed by applying the Interval Newton method~\cite{moore2009intervalanalysis,ishii2016intervalmonitoring,berz2001verifiedinversiontm} to $H^{(k)}_p$ (using its derivative $\frac{\mathrm d}{\mathrm d t}H_p^{(k)}$ which may be derived by Taylor model differentiation~\cite{berz1998tm}), and are guaranteed to enclose the roots of $p$.
Then $\rho$ must have a consistent boolean value in between these roots, which we may sample by performing interval evaluation of $H^{(k)}_p$~(see \cref{fig:monitoring-atomic}).

\subsection{Efficient Monitoring of Composed Taylor Models}

\label{sec:efficient-taylor-model-monitoring}

The method described in \cref{sec:signals-for-atomics} relies on being able to efficiently compute the interval function $H_{p}^{(k)}$ defined as a symbolic composition of Taylor models (\cref{fig:p-taylor-model}).
This is potentially very expensive since the composition involves symbolic operations on high-order polynomials and a Flowpipe may consist of thousands of time steps,
each requiring a separate composition. 

However, since we only need to deduce the signal for the atomic proposition,
rather than the exact function value at each point,
it will often be sufficient to inexpensively over-approximate the range of \cref{fig:p-taylor-model} over the current time step via interval arithmetic,
which we do by replacing some of the Taylor model compositions (denoted $\square$) with functional compositions (denoted $\circ$).
Hence, we use the following adaptive algorithm:
\begin{itemize}
    \item Perform the interval evaluation stepwise using interval arithmetic to check if
    \[
        0 \in \operatorname{range}{\left[
        p \circ
        \left(
            \mathbf q_{\text{post}}^{(k)},\,
            \mathbf I_{\text{post}}^{(k)}
        \right)
        \circ
        \left(
            \mathbf q_{\text{pre}}^{(k)},\,
            \mathbf I_{\text{pre}}^{(k)}
        \right)
        \right]}
    \]
    
    \item If so, perform one stage of symbolic composition and check if
    \[
        0 \in \operatorname{range}{\left[
        p \circ
        \left(
            \mathbf q_{\text{post}}^{(k)},\,
            \mathbf I_{\text{post}}^{(k)}
        \right)
        \fpcomp
        \left(
            \mathbf q_{\text{pre}}^{(k)},\,
            \mathbf I_{\text{pre}}^{(k)}
        \right)
        \right]}
    \]
    
    \item If the result is still ambiguous, perform full symbolic composition of $G_p^{(k)}$ for the current time step and apply root finding.
\end{itemize}

Hence, we are able to generate a precise signal for an atomic proposition over the whole time domain, whilst only performing symbolic Taylor model composition and root finding on demand where necessary to disambiguate the result of the signal (i.e. near the roots of the atomic proposition).
We may additionally skip the composition of the preconditioned Taylor model on dimensions which do not correspond to any variable of $p$.

This method may, however, still spend effort trying to determine the truth value of the signal of an atomic proposition in regions of time which are not crucial to the truth of the overall signal; this issue is addressed in the next section with the introduction of \emph{masks}.

\section{Masks}
\label{sec:masks}

In this section we introduce \emph{masks} which allow us to direct the monitoring process to certain time-regions on the flowpipe.
We then see how the mask required for each proposition may be constructed in a top-down manner, taking into account the context of an atomic proposition in the overall STL monitoring process (\cref{sec:monitoring-contexts,sec:computing-masks}).
Once we have an appropriate mask, in~\cref{sec:monitoring-under-masks} we see how to reduce the cost of monitoring an atomic proposition by avoiding work associated within time points outside of the mask.

\subsection{Basic Notions}
\label{sec:mask-notions}

Firstly we introduce \emph{masks} as follows:
\begin{definition}
    A \emph{mask} is a finite sequence $m = (I_j)_j$ of disjoint intervals~$I_j$. We refer to these intervals as the \emph{regions of interest under the mask $m$}.
\end{definition}
We can interpret a mask $m$ as a boolean signal $m:[0, \infty) \to \{\true, \false\}$ such that $m(x) = \true$ iff $x\in\bigcup_j I_j$.

Such a mask represents the region of time for which we wish to monitor a given proposition.
Since for soundness of monitoring we only need to over-approximate these regions of interest, in a practical implementation we may restrict ourselves to masks whose components are all closed intervals $I_j \triangleq [a_j, b_j]$ (using e.g. floating point endpoints) and consistently round outwards. We will however sometimes use other types of interval endpoints in what follows in order to state crisp results. 

We can apply a mask to an existing signal, erasing any truth values that lie outside the mask.
\begin{definition}
	Given a signal~$s$ and a mask $m$, the \emph{masked signal of $s$ by $m$} is the signal~$s|_m$ defined as
	\[
		s|_m(t) \;\triangleq\; \begin{cases}
			s(t) & \text{if $m(t) = \true$} \\
			\unknown & \text{otherwise.}
		\end{cases}
	\]
\end{definition}

\subsubsection*{Examples of Masking}

Before laying out rules for using and computing masks, we will illustrate their use in two different examples, demonstrating the importance of the temporal and logical context of a proposition within a STL formula.
\begin{example}
	\label{ex:F-mask}

	Suppose we want to monitor the property
	\(
		\varphi \triangleq \F{5}{6} \psi
	\)
	for $2$ seconds (that is, over the time-domain $I \triangleq [0, 2]$). This would naively require computing a signal for~$\psi$ over $8$ seconds, despite the fact that $\varphi$ only looks at the behaviour of~$\psi$ between $5$ and~$6$ seconds in the future --- that is, within the absolute time-domain
	\[
		I + [5, 6] = [0, 2] + [5, 6] = [5, 8].
	\]
	This means that in checking~$\varphi$ it is sufficient to compute a signal for~$\psi$ under the mask $m \triangleq ([5, 8])$, allowing us to ignore more than half of the time-domain.
\end{example}

\begin{example}
	\label{ex:or-mask}

	Suppose we want to monitor the property 
	\(
		\varphi \triangleq \psi \vee \sigma
	\)
	for $5$ seconds (that is, over the time-domain $I \triangleq [0, 5]$).
	This would normally require computing signals for $\psi$ and~$\sigma$ over the whole time domain~$I$.
	However, if we have already computed a signal for~$\psi$ such as
	\[
		s \triangleq (([0,1], \false), ([2, 4], \true))
	\]
	then
	it is evident that computing a signal for~$\phi$ only depends on the truth value of~$\sigma$ on the intervals $[0, 2)$ and $(4, 5]$.
	It thus suffices to compute a signal for~$\sigma$ under the mask
	\(
		m \triangleq ([0, 2), (4, 5]).
	\)
	This demonstrates how masks can enable a form of \emph{temporal short-circuiting}.
\end{example}

Whilst in both of the above examples the required masks are quite simple, for general propositions they quickly become much more complex depending on what signals we have already computed for parts of the property (as in \cref{ex:or-mask}) or the position of the current proposition in a larger property. Later in this section we will see how the reasoning in these two examples can be generalised and combined to build up masks for arbitrary properties in a compositional manner. 

\subsubsection{Operations on Masks}

We next need to define the operations with which masks can be build. Firstly, masks inherit all of the normal logical operations on boolean signals~\cite{maler2004monitoring}. In particular, given masks $m_I = (I_k)_{k}$ and $m_J = (J_l)_{l}$ we have that
\(
	m_I \wedge m_J = (I_k \cap J_l)_{l,k},
\)
and we write the negation of a mask $m$ as $\neg m$.

We will also need the temporal operators $\PI{J}m$ \emph{(past)} and $\HI{J}m$ \emph{(historically)} defined on masks by,
\begin{definition}
	\label{dfn:mask-shift} 
	Given a mask $m_I = (I_k)_k$ and an interval $J = [a, b]$, the \emph{past mask} is defined by,
	$$
		\PI{J}{m_I} \triangleq \left(I_k + J\right)_k
	$$

	whilst the \emph{historically mask} is defined by, 
	$$
		\HI{J}{m_I} \triangleq \neg \PI{J}{(\neg m)} = \left((I_k + a) \cap (I_k + b)\right)_k
		.
	$$
\end{definition}

\subsection{Monitoring Contexts}

\label{sec:monitoring-contexts}

Before we specify how the masks for the monitoring algorithm should be computed, we must first formalise what is required of a mask for it to be used at a given stage of the monitoring algorithm. 
This motivates us to define~\emph{contexts} which capture our existing knowledge at each recursive step of the monitoring algorithm, by recording the position of an argument~$\psi$ within the STL operator currently being monitored and the signals~$s$ we have already computed for any other arguments.
\begin{definition}
	A \emph{monitoring context} is defined similarly to a STL formula except with subformulae replaced by concrete signals~$s$ and, in exactly one place, a \emph{hole}~$[\cdot]$.
	That is, a monitoring context is defined according to the grammar
	\begin{align*}
		\mathcal C([\cdot])	
		\bnfdef [\cdot] &
		\bnfbar s \vee \mathcal C ([\cdot])
		\bnfbar s \wedge \mathcal C ([\cdot])
		\bnfbar \neg\,\mathcal C([\cdot]) \\
		& \bnfbar \FI{I}(\mathcal C([\cdot]))
		\bnfbar \GI{I}(\mathcal C([\cdot]))
		\bnfbar s \UI{I} \mathcal C([\cdot])\;\text{.}
	\end{align*}
	A monitoring context $\mathcal C([\cdot])$ is a \emph{monitoring context of the subformula~$\psi$ of a STL formula~$\phi$}, if $\mathcal C([\cdot])$ has the same structure as~$\phi$ except that, in the place of each atomic proposition $\rho$ of~$\phi$, $\mathcal C([\cdot])$ has a signal~$s_\rho$ that is a signal for~$\rho$, and the hole $[\cdot]$ in place of~$\psi$.
\end{definition}
Given a signal~$s$, we can evaluate a monitoring context $\mathcal C([\cdot])$ to give a signal~$\mathcal C(s)$ by substituting $s$ in the place of the hole~$[\cdot]$ and following the usual rules for combining signals. This means that a monitoring context captures how the signal for the overall formula depends on the signal for the proposition which remains to be monitored.

We are now able to define when a mask is sufficient for monitoring in a context.
\begin{definition}
	A mask $m$ is \emph{sufficient for a monitoring context $\mathcal C([\cdot])$ under mask $n$}, if for any signal $s$ we have that
	\[
		\mathcal C(s)|_n =  \mathcal C(s|_m)|_n.
	\]
\end{definition}
That is, a mask is sufficient if signals masked by it are \emph{just as good} as unmasked signals for monitoring the overall formula.

We also wish to know when a mask is as small as possible for monitoring in a given context.
\begin{definition}
	A mask $m$ is the \emph{optimal mask in context $\mathcal C([\cdot])$ under mask $n$} if it is the smallest sufficient mask in context $\mathcal C([\cdot])$ under mask $n$ with respect to the pointwise logical ordering~$\leq$.
\end{definition}
It follows directly that the mask defined above is unique (for a given context and overall mask), allowing us to talk about \emph{the mask} for a given context.

\subsection{Monitoring Under a Mask: Complex Propositions}
\label{sec:computing-masks}

We are now ready to detail how our masked monitoring algorithm deals with complex propositions,
by introducing suitable masks for each temporal and logical context.
In each case we prove that these masks are sufficient (and optimal) for the relevant context, which collectively shows the correctness of masked monitoring. 

\subsubsection{Negation}

Suppose we want to monitor a negation $\neg \phi$ under mask $m$, then this is equivalent to monitoring $\phi$ under mask $m$ and then negating the resulting signal.

\begin{proposition}
	The mask $m$ is itself sufficient and optimal for monitoring under $m$ in the monitoring context
	\[
		\mathcal C([\cdot]) = \neg [\cdot].
	\]
\end{proposition}
\begin{proof}
	~
	\paragraph*{Sufficiency}
	Take any signal~$s$ and any time $t \in [0, \infty)$. Then if $m(t)$,
	\[
		  (\neg (s|_m))|_m(t)
		= (\neg (s|_m))(t)
		= \neg (s|_m)(t)
		= \neg s(t)
		= (\neg s)|_m(t)
	\]	
	and if $\neg m(t)$,
	\[
		  (\neg (s|_m))|_m(t)
		= \unknown
		= (\neg s)|_m(t)
	\]

	\paragraph*{Optimality}
	Let $m'$ be any sufficient signal.
	Take the signal $s(t) \equiv \false$ and choose any time $t \in [0, \infty)$ such that $m(t) = \true$.
	Then, by the sufficiency of $m'$ we must have
	\[
		\true
		= \neg s(t)
		= (\neg (s|_{m'}))(t)
		= \begin{cases}
			\true & \text{if $m'(t) = \false$} \\
			\unknown & \text{otherwise}
		\end{cases}
	\]
	and so we must have $m'(t) = \true$ showing $m \Rightarrow m'$.
\end{proof}

\subsubsection{Eventually \texorpdfstring{($\F{a}{b} \phi$)}{} and Globally \texorpdfstring{($\G{a}{b} \phi$)}{}}

Suppose we want to monitor the property $\F{a}{b} \phi$ or $\G{a}{b} \phi$, under mask $m = (I_j)_j$.
In this case we should monitor $\phi$ under the past mask
$$
	\PI{[a, b]} m = (I_j + [a, b])_j.
$$
because the truth of $\phi$ at time $t$ could determine the truth of either $\F{a}{b}\phi$ or $\G{a}{b} \phi$ at any point between $a$ and $b$ seconds ago (in the former case by witnessing its truth, and in the latter case, by witnessing its falsehood) --- this generalises the reasoning given in \cref{ex:F-mask}.

\begin{proposition}
	Given a context
	\begin{alignat*}{3}
		\mathcal C([\cdot]) = \F{a}{b} [\cdot]
		& \qquad\text{or}\qquad &
		\mathcal C([\cdot]) = \G{a}{b} [\cdot]
	\end{alignat*}
	under the overall mask $m$,
	in each case the mask $\Pab{a}{b} m$ is sufficient and optimal for $\mathcal C([\cdot])$.
\end{proposition}
\begin{proof}
	Here we just prove sufficiency and optimality for $\mathcal C([\cdot]) = \F{a}{b} [\cdot]$ the results for $\mathcal C([\cdot]) = \G{a}{b} [\cdot]$ follow since $\G{a}{b} \phi \equiv \neg \F{a}{b} \neg \phi$.

	\paragraph*{Sufficiency}
	For sufficiency we need to show that
	\[
		\mathcal C{\left(s|_{\Pab{a}{b} m}\right)}(t)
		= \F{a}{b}\!{\left(s|_{\Pab{a}{b} m}\right)}(t)
		= \F{a}{b}(s)(t)
		= \mathcal C(s)(t)
	\]
	for any three-valued signal $s$ and time point $t$ such that $m(t) = \true$. We do this by showing that $\Pab{a}{b}m(t') = \true$ and hence $s|_{\Pab{a}{b} m}(t') = s(t')$ at each of the future time points $t' \in t + [a, b]$ to which both of the above $\F{a}{b}$~operators refer.
	This holds by contrapositive since if we had some $t' \in t + [a, b]$ for which $\Pab{a}{b} = \false$, then we would have $m(t'') = \false$ for all $t'' \in t' - [a, b]$ and, in particular, $m(t) = \false$.

	\paragraph*{Optimality}
	Suppose $m'$ is any mask such that $m' < \Pab{a}{b} m$. Then we have some $t_0$ for which $m' (t_0)$ whilst $\Pab{a}{b} m (t_0)$ and hence we must have some $t_0' \in t_0 - [a, b]$ such that $m(t') = \true$. But then if we take the signal
	\[
		s(t) = \begin{cases}
			\true & \text{if $t = t_0$} \\
			\unknown & \text{otherwise}
		\end{cases}
	\]
	we see that
	\(
		\F{a}{b} s(t_0') = \true
	\)
	whilst
	\(
		\F{a}{b} s(t_0') = \unknown
	\)
	and hence $m'$ is not sufficient, proving the optimality of $\Pab{a}{b} m$.
\end{proof}

\subsubsection{Disjunctions and Conjunctions}

Suppose we want to monitor a disjunction $\phi \vee \psi$ under mask $m$.
We should first monitor $\phi$ under the mask $m$ to give the signal $s$. Then, generalising~\cref{ex:or-mask}, we can use the signal~$s$ to generate a mask $m_{s}^{\vee}$, the \emph{or-mask of $s$}.
\begin{definition}
	Given a three-valued signal~$s = (I_j, s_j)_j$, the \emph{or-mask of $s$} is the mask $m_s^{\vee}$ defined by $m_s^{\vee}(t) = \true$ iff $s(t) \in \{\false, \unknown\}$ so
	$$
		m_s^{\vee} = \bigwedge_{s_j = \true} {m_j}
	$$
	where $m_j \triangleq \left(C_j^{(\ell)}, C_j^{(u)} \right)$ is the mask consisting of the two interval complements $C_j^{(\ell)}, C_j^{(u)}$ of $I_j$ in $[0, \infty)$.
\end{definition}
If this mask turns out to be empty (i.e. if $s(t) = \false = m_s^{\vee}(t)$ for all $x \in m$), then we can stop and conclude $s$ is a signal for $\phi \vee \psi$ under $m$.
Otherwise, we monitor $\psi$ under the mask $m_{s}^{\vee}$ giving a signal~$w$, and hence the signal~$s \vee w$ for $\phi \vee \psi$ under $m$.

We see that the mask~$m_{s}^{\vee}$ is optimal and sufficient for the context $\mathcal C([\cdot]) = s \vee [\cdot]$.
\begin{proposition}
	Given a monitoring context,
	\(
		\mathcal C([\cdot]) = s \wedge [\cdot]
	\)
	the and-signal $m^{\wedge}_s$ is sufficient and optimal for this context under the mask $m$.
\end{proposition}
\begin{proof}
	~
	\paragraph*{Sufficiency}
	Take any signal~$w$ and any time $t \in [0, \infty)$ such that $m(t) = \true$. Then if $s(t) = \false$,
	\[
		\mathcal C(w|_{m^{\wedge}_s})(t)
		= \false \wedge w|_{m^{\wedge}_s}(t)
		= \false
		= \false \wedge w(t)
		= \mathcal C(w)(t)
	\]
	whilst if $s(t) \in \{\true, \unknown\}$, then $m^{\wedge}_s(t) = \true$ and hence 
	\[
		\mathcal C{\left(w|_{m^{\wedge}_s}\right)}(t)
		= s(t) \wedge w|_{m^{\wedge}_s}(t)
		= s(t) \wedge w(t)
		= \mathcal C(w)(t)
	\]
	showing $m^{\wedge}_s$ is sufficient.

	\paragraph*{Optimality}
	Let $m'$ be any mask such that $m' < m^{\wedge}_s$. Then there must be some time $t_0$ such that $m'(t_0) = \false$ and $m^{\wedge}_s(t_0) = \true$ and hence $s(t_0) \in \{\true, \unknown\}$. But then if we pick the signal $w(t) \equiv \false$, we see that
	\[
		\mathcal C(w|_{m'})
		= s(t_0) \wedge (s|_{m'})(t_0)
		= s(t_0) \wedge \unknown
		= \unknown
	\]
	whilst 
	\[
		\mathcal C{\left(w|_{m^{\wedge}_s}\right)}(t)
		= s(t_0) \wedge w|_{m^{\wedge}_s}(t_0)
		= s(t_0) \wedge \false
		= \false
	\]
	and hence $m'$ is not sufficient, proving that $m^{\wedge}_s$ must be optimal.
\end{proof}

We treat conjunctions similarly, and can see that the that \emph{and-mask}~$m^{\wedge}_s$ defined by $m^{\wedge}_s(t) = \bigwedge_{s_j = \false} {m_j}$ is an optimal and sufficient mask for conjunctions $\mathcal C([\cdot]) = s \wedge [\cdot]$.

\subsubsection{Until \texorpdfstring{($\phi \U{a}{b} \psi$)}{}}

Finally, suppose we wish to monitor the signal for the property $\phi \U{a}{b} \psi$ under the mask $m$.
As in~\cref{sec:signals}, we will compute the signal for $\phi \U{a}{b} \psi$ based on signals for $\phi$ and $\psi$ using \cref{eq:until-decomposition},
however we now need to monitor $\phi$ and $\psi$ under appropriate masks.
We start by monitoring~$\phi$ under the mask $m \vee \Pab{a}{b} m$ (taking into account the two places in which it appears in \cref{eq:until-decomposition}).
Then we could find a suitable mask for $\psi$ by applying the above rules for $\vee$, $\wedge$, and $\F{a}{b}$ to \cref{eq:until-decomposition}.
However, it turns out that this mask may be computed directly using the historically operator. 

For this we first need the case of unitary masks,
\begin{restatable}{proposition}{Hunitaryequivalence}
	\label{prop:H-unitary-equivalence}
	Given an unitary mask $m$ we have that
	\[
		\HI{[0, a]} m = m \wedge \PI{[a, b]} m.
	\]
\end{restatable}
\begin{proof}
	Given in \cref{appendix:until-mask-proof} (\cref{prop:H-unitary-equivalence}).
\end{proof}

And we can see historically distributes over disjoint unitary masks.
\begin{restatable}{proposition}{Hdecompositionstep}
	\label{prop:H-decomposition-step}
	Given disjoint, closed unitary masks $m, n$ and an interval $J = [a, b]$, we have that
	\[
		\HI{J}(s \vee w) = \HI{J} s \vee \HI{J} w .
	\]
\end{restatable}
\begin{proof}
	Given in \cref{appendix:until-mask-proof} (\cref{prop:H-decomposition-step}).
\end{proof}

Together these give us the mask for the until operator.
\begin{restatable}{proposition}{untilmask}
    \label{prop:until-mask}
	The mask
	\[
		m^{\UI{a}}_s \triangleq \Hab{0}{a} \left(m^{\wedge}_s\right)
	\]
	is optimal and sufficient for monitoring context $\mathcal C([\cdot]) = s \U{a}{b} [\cdot]$.
\end{restatable}
\begin{proof}
	Given in \cref{appendix:until-mask-proof} (\cref{prop:until-mask}).
\end{proof}

\subsection{Monitoring Under a Mask: Atomic Propositions}
\label{sec:monitoring-under-masks}

Once we have determined a mask~$m = (I_j)_j$ for a given atomic proposition~$\rho$ given its context in the monitoring process, we then aim to directly monitor a signal~$s$ for~$\phi$ \emph{under the mask~$m$}.
This means that we only care about the value of $s(t)$ at time points $t$ for which $m(t)$ is true, and so can increase the efficiency of monitoring by avoiding work associated with time points outside of the mask. 
Whilst there is no way to save Flow* from having to generate the flowpipes for these time points (since they may be required for determining the future evolution of the system), we can avoid the effort associated with every subsequent step of the monitoring process.

We do this by modifying how we carry out monitoring of $\rho$ (via $H^{(k)}_p$) on each flowpipe segment (\cref{sec:efficient-taylor-model-monitoring}) over its associated time domain $T_k = [t_k, t_{k+1}]$ as follows:
\begin{itemize}
    \item if $m \wedge (T_k) = \varnothing$ then we set $s(t) = \unknown$ for all $t \in T_k$ and avoid monitoring over this segment;
    
	\item otherwise, we restrict the time domain to the interval $T_k^\prime = \bigcup_j T_k \cap I_j$ and apply the normal monitoring process.
\end{itemize}

This immediately saves us from performing root finding on regions outside of the mask. Additionally, since we have already seen how the symbolic composition of the preconditioned Taylor model flowpipe and between the flowpipe and the atomic propositions may be performed on demand, these expensive operations may also be avoided outside of the mask. Thus masks allow us to direct the monitoring process for each atomic proposition based on its context within a wider STL formula.

\section{Demonstration and Performance Analysis}
\label{sec:evaluation}

\begin{figure}[t]
\centering
\noindent\begin{minipage}{0.48\textwidth}
    \centering
    \includegraphics[width=\textwidth]{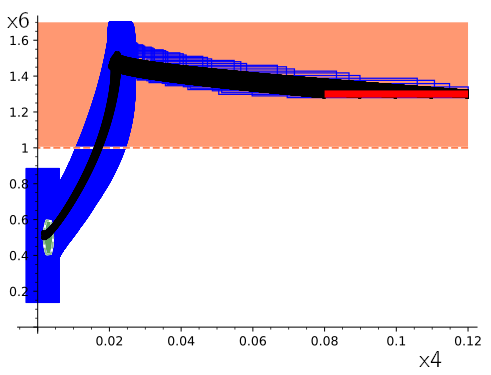}
    \vspace{-6ex}
    \captionof{figure}{
        An interval over-approximation of the Flow* flowpipe at each time step is illustrated in \fcolorbox{blue}{white}{blue}, numerical trajectories for different initial conditions in \fcolorbox{black}{white}{black}, initial conditions in \fcolorbox{red}{white}{red}, and the regions involved in properties P and Q are in \colorbox{Pcol}{orange} and \colorbox{Qcol}{green} respectively.
    }
    \label{fig:genetic-oscillator}
\end{minipage}%
\hfill%
\begin{minipage}{0.485\linewidth}
    \subfloat[Functional composition\label{fig:Qfunccomp}]{%
      \includegraphics[width=\linewidth]{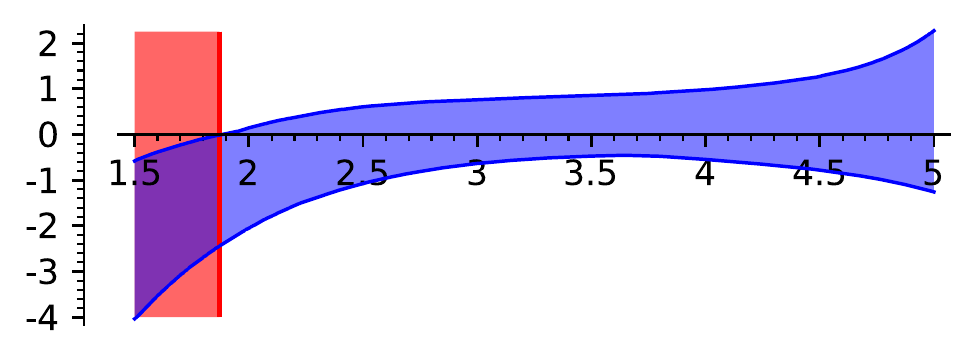}%
    }%
    \newline%
    \subfloat[Symbolic composition\label{fig:Qsymbcomp}]{%
      \includegraphics[width=\linewidth]{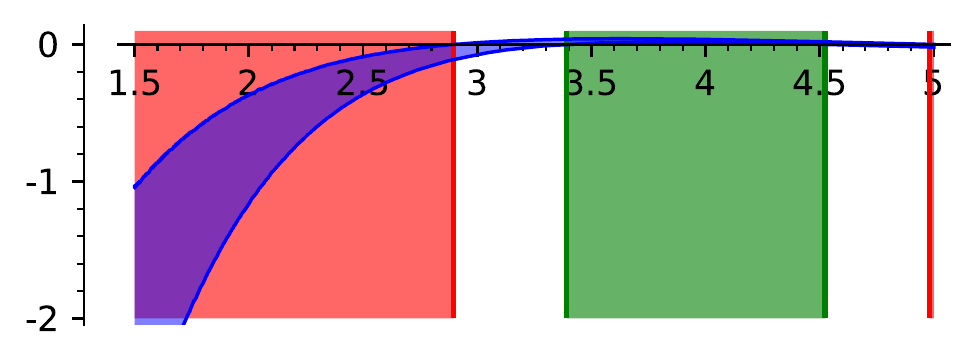}%
    }
    \newline%
    \subfloat[Mask for $Q$ in context\label{fig:Qmask}]{%
      \hspace{0.05\linewidth}\includegraphics[width=0.95\linewidth]{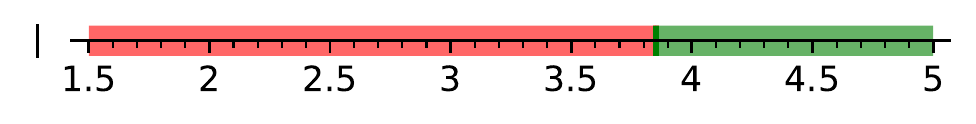}%
    }%
    
    \captionof{figure}{Monitoring $Q$.}
\end{minipage}
\end{figure}
We have implemented the monitoring techniques discussed in this report as a Python library with Cython~\cite{behnel2010cython} code implementing interval and Taylor model operations by interfacing with Flow*'s internal C++ libraries.
In this section we will use this implementation to demonstrate the application of our method to verifying STL properties of a known challenging continuous system,
the 9-dimensional genetic oscillator~\cite{vilar2002geneticoscillator,chen2016decomposed} involving non-linear (polynomial) ODEs over the variables
\(
  x_1, \ldots, x_9
\)
and an uncertain box of interval initial conditions $x_i \in [a_i, b_i]$
(given in full along with other details of our benchmarking approach in \cref{appendix:benchmark-details}).
The evolution of the variables $x_4$ and $x_6$ over 5 seconds is shown in \cref{fig:genetic-oscillator} which includes numerical traces from a sample of many different fixed initial conditions alongside a coarse interval over-approximation of a Flow* flowpipe covering the whole box of uncertain initial conditions.
We can describe the temporal behaviour of this system much more precisely with STL properties such as
\(
    \phi
    \triangleq
    \G{0}{1}\!{\left(P \vee \G{3}{3.5}\!{(Q)}\right)}
\)
in which we have polynomial atomic propositions
\(
    P \triangleq x_6 - 1 > 0
\)
and
\(
    Q
    \triangleq
    0.032
    -
    125^2 (x_4 - 0.003)^2 
    -
    3 (x_6 - 0.5)^2
    > 0
    .
\)
The property $\phi$ states that at any point within the first second,
the system will either remain within the half-space $P$ or,
at any point between $3$ and $3.5$ seconds in the future will be within the elliptical region $Q$.

\begin{figure}
    \centering
    \includegraphics[width=\linewidth]{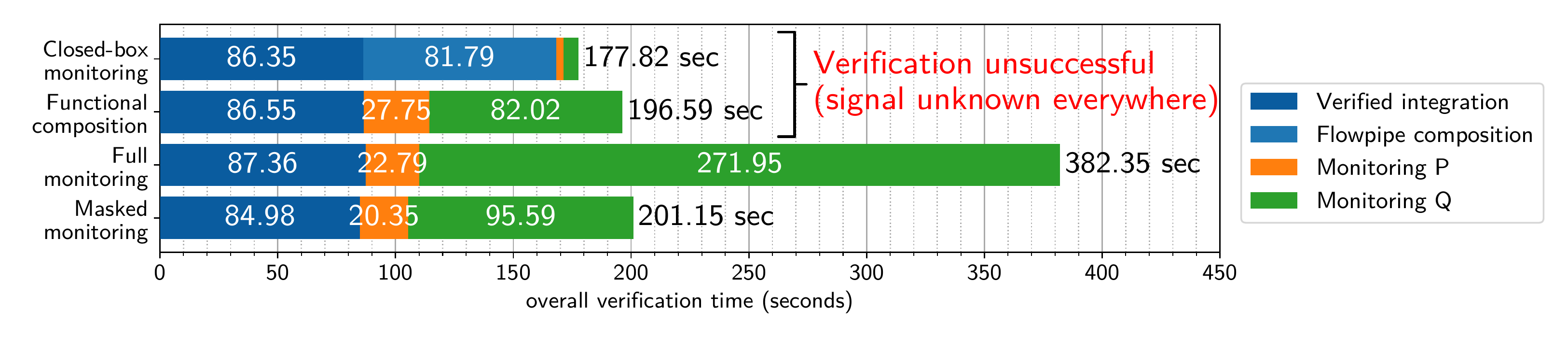}
    \caption{Combined Flow* verified integration and STL monitoring times in seconds, showing the cost of each stage for a number of variants of our monitoring algorithm.}
    \label{fig:benchmark-bar-chart}
\end{figure}

In \cref{fig:benchmark-bar-chart} we break down the time taken to monitor~$\phi$ for $0.5$ seconds using a number of variants of our monitoring algorithm in order to evaluate the impact of each of its elements on monitoring cost and precision.
First we consider the closed box monitoring approach where we first run Flow* to perform verified integration and flowpipe composition,
before using interval analysis and functional composition to monitor $\phi$ over the entire flowpipe.
Whilst the monitoring cost for the propositions $P$ and $Q$ is very small in comparison to the time it took Flow* to perform verified integration,
the flowpipe composition stage is more expensive and takes almost as long as the verified integration itself.
Next we monitor $\phi$ in the same way, but perform the flowpipe composition on demand as described in \cref{sec:efficient-taylor-model-monitoring}.
We see that if we just monitor the simple atomic proposition $P$ we save most of the cost of flowpipe composition, although once we also monitor $Q$ we need to pay the full cost.
These two methods also do not yield sufficient precision to verify $\phi$, both producing a useless signal which is unknown everywhere.
This imprecision can be seen in \cref{fig:Qfunccomp} which shows the result of monitoring the complex polynomial atomic proposition $Q$ over the flowpipe using functional composition and the corresponding signal.

In order to produce a useful signal for $\phi$ we need to run our full monitoring algorithm, permitting symbolic composition at each stage.
Whilst the monitoring cost for the simple proposition $P$ is similar to before,
the cost for the complex proposition $Q$ is significantly higher.
This, however, now gives a much more precise signal for $Q$ as shown in \cref{fig:Qsymbcomp}.
This means we now get the overall signal $s = (([0,0.0237], \true))$ for $\phi$, allowing us to verify that $\phi$ is true at time $0$ and that $P \vee \G{3}{3.5}(Q)$ holds for at least the first $1.0237$ seconds.
Finally, we reran our monitoring algorithm but monitored each atomic proposition under appropriate masks.
For example, $Q$ is monitored under the mask shown in \cref{fig:Qmask}.
This produced the same overall signal as the full unmasked monitoring algorithm but reduced the monitoring time for $Q$ by 65\%.

\section{Conclusion and Future Work}
\label{sec:conclusion}

In this report we explored a symbolic algorithm for monitoring STL properties over Taylor model flowpipes via three-valued signals and introduced masking to direct the monitoring process to time regions critical to the property at hand.
We saw that,
whilst direct integration with the symbolic structure of flowpipes can add some overhead for complex propositions,
it significantly increases the precision of the results generated and is sometimes necessary to produce definitive results at all.
We have also seen that masking can have an important impact on reducing this cost,
by avoiding the need to carry out symbolic operations for regions of time not necessary for the overall result of the monitoring algorithm.

Our current method relies on interval analysis to evaluate flowpipes over the whole set of uncertain initial conditions,
whereas Flow*'s flowpipes in fact track the functional dependency of each system variable on the initial condition~\cite{berz1998tm}.
In future we intend to use this additional information to extend our algorithm to produce spatio-temporal signals exploring refinements of the initial conditions and introduce spatio-temporal masks to enable \emph{spatio-temporal short-circuiting}.

\subsubsection*{Acknowledgements}

We would like to thank Paul Jackson for many useful discussions of Taylor models and Flow*, Chris Banks for providing feedback on our approach, Kristjan Liiva for useful discussions of Flow*'s code and for his patches to Flow* for native floating point arithmetic, and Jos Gibbons and Juliet Cooke for providing feedback on drafts of this report. Thanks also go to our anonymous reviewers for helpful feedback, and for pointing out the connections of our work to~\cite{deshmukh2017robustintervalstl}.

This work was supported by the Engineering and Physical Sciences Research Council (grant EP/L01503X/1), EPSRC Centre for Doctoral Training in Pervasive Parallelism at the University of Edinburgh, School of Informatics.

\printbibliography

\begin{subappendices}
    \renewcommand{\thesection}{\Alph{section}}%
    \crefalias{section}{appendix}
    \crefalias{subsection}{appendix}

\section{Proofs for Until Mask}

\label{appendix:until-mask-proof}

For the sufficiency and optimality proofs of the until mask, we first need to consider the case of unitary masks,
\Hunitaryequivalence*
\begin{proof}
	Suppose $\true = m \wedge \PI{[a, b]}(m)(t)$. Then $\true = s(t)$ and for some $t' \in t + [a, b]$, $s(t') = \true$. But then for any $t'' \in t + [0, a]$, we must have $s(t'') = \true$ since $m$ is unitary, showing $\HI{[0, a]} s (t) = \true$.

	Conversely, suppose $\HI{[0, a]} s (t) = \true$. The for all $t' \in t + [0, a]$ we have that $s(t) = \true$. In particular $s(t) = \true$ and $s(a + t) = \true$ showing that $s \wedge \PI{[a, b]} s = \true$.
\end{proof}

Next we can see historically distributes over disjoint unitary masks.
\Hdecompositionstep*
\begin{proof}
	Take any time $t$. If $\HI{J}s(t)$ or $\HI{J}w(t)$ then in either case we clearly have that $\HI{J}(s \vee w)(t)$.
	
	Conversely, if $\HI{J}(s \vee w)(t)$ then for any $t' \in J$, $s(t') = \true$ or $w(t') = \true$. If neither of these signals is true both endpoints of $t + J$ (in which case we would have $\HI{J} s \vee \HI{J} w (t') = \true$ by unitarity) we can suppose w.l.o.g. that $s(t + a) = \true$ and $w(t + b) = \true$. Then let
	\begin{alignat*}{8}
		l && {}={} && \sup \bigl\{ t' \in t + J \mathrel: s(t') = \true \bigr\}
		&& \quad \text{and} \quad &&
		u && {}={} && \inf \bigl\{ t' \in t + J \mathrel: w(t') = \true \bigr\}.
	\end{alignat*}
	Suppose $l \leq u$. Then if we let $m = \frac{l + u}{2}$, we must have either $s(m) = \true$ or $w(m) = \true$. W.l.o.g. assuming the former, we must have $l = m = \frac{l + u}{2} \leq u$ and hence $l = u$, a contradiction. Therefore, we must have $l > u$. But then we have $s = w$ by unitarity and disjointness, and hence $\HI{J}(s \vee w)(t) = \HI{J} s (t) = \HI{J} s \vee \HI{J} w(t) = \true$.
\end{proof}

\begin{proposition}
	If $m$ is any mask such that $m = \bigvee_j m_j$ ($m_j$ disjoint closed  unitary signals) then
	\[
		\HI{[0, a]} m = \bigvee_j m_j \wedge \PI{[a, b]} m_j .
	\]
\end{proposition}
\begin{proof}
	Follows by \cref{prop:H-unitary-equivalence} and \cref{prop:H-decomposition-step}. 
\end{proof}

We can now use the until decomposition to prove the overall result.
\untilmask*
\begin{proof}
	For both proofs we will use the decomposition of $w$ into disjoint closed components
	\[
		w = \bigvee_{j,k} w_{j, k}.
	\]
	~
	\paragraph{Sufficiency}
	For any signal~$s$ we have
	\[
		w \U{a}{b} s|_{m}
		=
		\bigvee_{j,k} w_{j, k}
			\wedge \F{a}{b}\left(
				w_{j, k} \wedge s|_{m}
			\right)
		.
	\]
	Take any $j, k$ and $t \in [0, \infty)$. If $w_{j,k}(t) = \false$ then
	\[
		w_{j, k}
			\wedge \F{a}{b}\left(
				w_{j, k} \wedge s|_{m}
			\right)(t)
		= \false
		= w_{j, k}
			\wedge \F{a}{b}\left(
				w_{j, k} \wedge s
			\right)(t)
	\]
	and we are done. If, on the other hand, $w_{j,k}(t) \in \{\true, \unknown\}$ and we take any $t' \in t + [a, b]$, such that
	\[
		m(t') = \Hab{0}{a} m^{\wedge}_w(t') = \false,
	\]
	then we have some $t'' \in t' - [0, a]$ such that
	\(
		m^{\wedge}_w(t'') = \false
	\)
	and hence
	\(
		w(t'') = \false
		.
	\)
	But then if we had 
	\(
		w_{j,k}(t') \in \{\true, \unknown\}
	\), since we know 
	\(
		w_{j,k}(t) \in \{\true, \unknown\}
	\)
	and $t'' \in [t, t']$, by unitarity we would have
	\(w_{j,k}(t'') \in  \{\true, \unknown\}\), a contradiction.
	Then, instead we must have $w_{j,k}(t') = \false$, and hence
	\[
		w_{j,k} \wedge s|_m (t')
		= \false
		= w_{j,k} \wedge s (t').
	\]
	Therefore, $w_{j,k} \wedge s|_m (t') = w_{j,k} \wedge s (t')$ for any $t' \in t + [a, b]$ and hence
	\[
		w \U{a}{b} s|_{m}
		= w \U{a}{b} s
	\]
	giving sufficiency.

	\paragraph{Optimality}
	Consider any mask $m'$ such that $m' < m$.
	Then there must be some $t_0 \geq 0$ such that $m(t_0) = \true$ whilst $m'(t_0) = \false$.
	Since
	\[ 
		m(t_0) = \Hab{0}{a} \left(m^{\wedge}_w\right)(t_0),
	\]
	we must have that $t_0 \geq a$ and
	for all $t' \in t_0$ we must have 
	\(
		m^{\wedge}_w(t') = \true
	\)
	and hence
	\(
		w(t') \in \{\true, \unknown\}.
	\)
	Then, by the nature of the decomposition of $w$, for any $j, k$, we must have either
	\begin{alignat}{3}
		\label{eq:decomposition_w_case_a}
		w_{j, k}(t') &\in \{\true, \unknown\}
			\quad&
			\text{for all $t' \in t_0 - [0, a]$} \\
		\intertext{or}
		\label{eq:decomposition_w_case_b}
		w_{j, k}(t') &= \false
			\quad&
			\text{for all $t' \in t_0 - [0, a]$}
	\end{alignat}

	Now, since we know $t_0 \geq a$, we may take $t \triangleq t_0 - a$ and define the signal~$s$ as
	\[
		s(t') \equiv \false.
	\]
	Then, in the case we have \cref{eq:decomposition_w_case_a}, we see
	\[
		w_{j,k}	\wedge \F{a}{b}\left(
			w_{j,k} \wedge s|_m
		\right)(t)
		= \false
	\]
	since $s|_m(t_0) = \false$ and for any $t' \in t + [a, b]$ for which $s_m(t') \neq \false$ we must have some $t'' \in [t' - a, t'] \subseteq [t, t']$ such that $w(t'') = w_{j,k}(t'') = \false$. In this case we have that, 
	\[
		w_{j,k}	\wedge \F{a}{b}\left(
			w_{j,k} \wedge s|_{m'}
		\right)(t)
		= \unknown
	\]
	since $s|_{m'}(t_0) = \unknown$ and $w_{j,k}(t') \in \{\true, \unknown\} $ for all $t' \in t_0 - [0, a] = [t, t_0]$.

	On the other hand, in the case we have \cref{eq:decomposition_w_case_b}, we have
	\[
		w_{j,k}	\wedge \F{a}{b}\left(w_{j,k} \wedge s|_m\right)(t)
		= \false
		= w_{j,k} \wedge \F{a}{b}\left(w_{j,k} \wedge s|_{m'}\right)(t),
	\]
	whence we conclude
	\(
		w \U{a}{b} s|_m (t) = \false
	\)
	whilst
	\(
		w \U{a}{b} s|_{m'} (t) = \unknown,
	\)
	showing $m'$ is not sufficient, and thus completing the proof.
\end{proof}

\section{Benchmark Details}

\label{appendix:benchmark-details}

The genetic oscillator~\cite{vilar2002geneticoscillator,chen2016decomposed} consists of the system of coupled ODEs:
\begin{alignat*}{2}
    \dot{x_1} & \triangleq 
    50 x_3 - 0.1 x_1 x_6 \\
    \dot{x_2} & \triangleq
    100 x_4 - x_1 x_2 \\
    \dot{x_3} & \triangleq
    0.1 x_1 x_6 - 50 x_3 \\
    \dot{x_4} & \triangleq
    x_2 x_6 - 100 x_4 \\
    \dot{x_5} & \triangleq
    5 x_3 + 0.5 x_1 - 10 x_5 \\
    \dot{x_6} & \triangleq
    50 x_5 + 50 x_3 + 100 x_4 - x_6 (0.1 x_1 + x_2 + 2 x_8 + 1) \\
    \dot{x_7} & \triangleq
    50 x_4 + 0.01 x_2 - 0.5 x_7 \\
    \dot{x_8} & \triangleq
    0.5 x_7 - 2 x_6 x_8 + x_9 - 0.2 x_8 \\
    \dot{x_9} & \triangleq
    2 x_6 x_8 - x_9
    \intertext{with uncertain initial conditions:}
    x_1(0) & \in \interval{0.98}{1.02} \\
    x_2(0) & \in \interval{1.28}{1.32} \\
    x_3(0) & \in \interval{0.08}{0.12} \\
    x_4(0) & \in \interval{0.08}{0.12} \\
    x_5(0) & \in \interval{0.08}{0.12} \\
    x_6(0) & \in \interval{1.28}{1.32} \\
    x_7(0) & \in \interval{2.48}{2.52} \\
    x_8(0) & \in \interval{0.58}{0.62} \\
    x_9(0) & \in \interval{1.28}{1.32}
\end{alignat*}

These benchmarks were carried out using Flow*'s polynomial symbolic remainder integration method with a remainder queue of $250$ time steps, a fixed Taylor model order of $4$, fixed step size of $0.003$, identity preconditioning, remainder estimation of $10^{-1}$ and a cutoff threshold of $10^{-6}$.

We took the times from averaging $30$ runs ensure consistency of results.
All benchmarks were run on an AMD Ryzen 7 3700U with 32 GB of RAM under Fedora 32 (Linux kernel 5.6.16 with Spectre v1-2 mitigations enabled).
Our implementation uses Flow* 2.1.0, patched to use native floating point arithmetic in favour of MPFR and to remove a memory leak.

\end{subappendices}

\end{document}